# ПАРАЛЛЕЛЬНЫЙ АЛГОРИТМ ДЛЯ АВТОМАТИЗИРОВАННОЙ РАЗМЕТКИ БОЛЬШИХ ВРЕМЕННЫХ РЯДОВ

© 2023 А.И. Гоглачев

*Южно-Уральский государственный университет*
*(454080 Челябинск, пр. им. В.И. Ленина, д. 76)*
*E-mail: goglachevai@susu.ru*

**Аннотация**: В данной статье представлен алгоритм PaSTiLa для автоматизированной разметки больших временных рядов на кластере с графическими процессорами. Метод автоматически подбирает значения длины сниппета на основе нового предложенного критерия и позволяет выполнять поиск шаблонов с высокой производительностью. Эксперименты показали высокую точность поиска шаблонов и преимущество метода по сравнению аналогами.
**Ключевые слова:** временной ряд, поиск шаблонов, матричный профиль, мера MPdist, параллельный алгоритм, графический процессор.

## 1 Введение

Поиск шаблонов временных рядов предполагает автоматизированный поиск типичных подпоследовательностей, которые выделяют различные активности, заданные указанным рядом. Шаблоны временного ряда позволяют кратко описать и визуализировать сенсорные данные и поэтому имеет широкий спектр практического применения: предиктивное техническое обслуживание в цифровой индустрии, умное управление системами жизнеобеспечения, мониторинг показателей функциональной диагностики организма человека, моделирование климата и др.

В нашей предыдущей работе [8] предложен параллельный алгоритм поиска шаблонов для графического процессора, названный PSF. В данной статье продолжается начатая работа. Недостатком разработанного ранее алгоритма является его чувствительность к значению параметра длины шаблона (сниппета). Неверный выбор этого значения может привести к значительному падению точности поиска шаблонов. В данной работе предлагается новый критерий по подбору наилучшего значения длины шаблона, который позволит исключить ошибки, связанные с неверным выбором значения параметра, и улучшит точность поиска шаблона. Для повышения производительности поиска шаблонов новый алгоритм разработан для работы на вычислительном кластере с узлами с графическими процессорами

Данная статья организована следующим образом. Раздел 2 содержит краткий обзор похожих по тематике работ. В разделе 3 приводится список, используемых в статье, определений и нотаций. Раздел 4 содержит подробное описание разработанного метода. Результаты вычислительных экспериментов, демонстрирующие качественные характеристики предложенного метода содержатся в Разделе 5. Заключение содержит сводку полученных результатов.

## 2 Обзор связанных работ

В [1] Кеог и др. описывают следующие свойства, которыми должны обладать типичные подпоследовательности, найденные во временном ряду: количественная оценка, разнообразие и независимость от предметной области. Количественная оценка предполагает возможность численного выражения сходства найденного шаблона с подпоследовательностью ряда и возможность упорядочивания шаблонов в соответствии с этой оценкой. Свойства разнообразия связано с уникальностью каждого найденного шаблона. Наконец, свойство независимости от предметной области предполагает, что алгоритм следует применять к временным рядам из любой предметной области без использования знаний и/или обучающих данных по этой области для достижения всех вышеперечисленных свойств. Кроме того, авторы обсуждают следующие очевидные подходы к обнаружению типичных подпоследовательностей временных рядов, которые частично или полностью не соответствуют вышеуказанным требованиям: мотивы [2], центроиды из кластеризации подпоследовательностей временных рядов [3] и шейплеты [7].

Лейтмотив [2] — это пара подпоследовательностей временного ряда, которые очень похожи друг на друга по выбранной мере схожести. В [2] Муин и др. предлагают эффективный алгоритм обнаружения мотивов на основе евклидова расстояния, который использует треугольное неравенство для отсечения неперспективных кандидатов. Однако концепция мотива не соответствует свойству количественной оценки, т.е. мы можем указать формы и расположение типичных найденных шаблонов, но не можем указать охват каждого шаблона.

В предметных областях, которые не связаны с временными рядами, суммирование набора объектов одинаковой структуры выполняется с помощью методов кластеризации. Кластеризация предполагает разделение заданного набора объектов на подмножества (кластеры) таким образом, чтобы объекты из одного кластера были существенно похожи друг на друга, в то время как объекты из разных кластеров существенно не похожи друг на друга по отношению к выбранной мере сходства [18]. Однако кластеризация всех подпоследовательностей временных рядов одинаковой длины бессмысленна при любом измерении расстояния или при любом алгоритме, как показано Кеогом и др. в [3].

В [7] Йе и др. представили концепцию шейплетов, которая предполагает, что подпоследовательности временных рядов предварительно классифицированы. Шейплет — это подпоследовательность, которая одновременно является наиболее похожей на большинство подпоследовательностей данного класса и наиболее непохожей на подпоследовательности, принадлежащие другим классам, в отношении выбранной меры сходства. Очевидно, концепция шейплета не зависит от предметной области.

В работе [1] Кеог и др. предложили концепцию сниппета временного ряда, которая отвечает всем вышеупомянутым свойствам. Неформально говоря, сниппет временного ряда — это подпоследовательность заданной длины, которая похожа на многие другие подпоследовательности временного ряда в отношении специально определенной меры схожести MPdist [6]. В тот же момент все подпоследовательности, аналогичные сниппету, могут быть точно указаны и подсчитаны. Набор сниппетов имеет значительно меньшую мощность, чем набор подпоследовательностей временного ряда заданной длины, и поэтому может быть использован для обобщения исходного временного ряда. При экспериментальной оценке обнаружение шаблонов на основе сниппетов показывает адекватные результаты для временных рядов из широкого спектра предметных областей.

## 3 Основные определения и нотации

Ниже приводятся обозначения и определения, используемые в данной статье терминов в соответствии с работами [1, 4, 5].

### 3.1 Временной ряд и подпоследовательность

*Временной ряд (time series)* $T$ представляет собой набор координат – векторов одинаковой длины. Каждая координата многомерного временного ряда представляет собой последовательность хронологически упорядоченных вещественных значений:

$$T^{(j)} = \left(t_1^{(j)}, \ldots, t_n^{(j)}\right), \ t_i^j \in \mathbb{R}, 1 \leq j \leq d. \tag{1}$$

Число $n$ обозначается $|T^j|$ и называется длиной ряда. Количество координат временного ряда обозначается как $d$.

*Элемент* $T_i$ представляет вектор, состоящий и $i$ элементов каждой координаты $T$:

$$T_i = \left(t_i^{(1)}, \ldots, t_i^{(d)}\right), \ t_i^j \in \mathbb{R}, 1 \leq i \leq n, 1 \leq j \leq d. \tag{2}$$

*Подпоследовательность (subsequence)* $T_{i,m}$ временного ряда $T$ представляет собой непрерывный промежуток из $m$ элементов, начиная с позиции $i$:

$$T_{i,m} = (t_i, \ldots, t_{i+m-1}), 1 \leq m \ll n, \ 1 \leq i \leq n - m + 1. \qquad (3)$$

Множество всех подпоследовательностей ряда $T$, имеющих длину $m$, обозначим как $S_T^m$, а мощность такого множества за $N$, $N = |S_T^m| = n - m + 1$.

## 3.2 Сниппеты (типичные подпоследовательности) временного ряда

Для упрощения изложения, в данном разделе под временным рядом $T$ будет пониматься одна из координат временного ряда $T^j$. В дальнейшей работе описанный ниже алгоритм применяется к каждой координате временного ряда независимо.

*Задача поиска типичных подпоследовательностей* временного ряда основана на концепции *сниппетов (snippet)*, которая предложена Кеогом и др. в работе [1] и уточняет понятие типичных подпоследовательностей временного ряда следующим образом. Каждый сниппет представляет собой один из сегментов временного ряда. Со сниппетом ассоциируются его ближайшие соседи – подпоследовательности ряда, имеющие ту же длину, что и сниппет, которые более похожи на данный сниппет, чем на другие сегменты. Для вычисления схожести подпоследовательностей используется специализированная мера схожести, основанная на евклидовом расстоянии. Сниппеты упорядочиваются по убыванию мощности множества своих ближайших соседей. Формальное определение сниппетов выглядит следующим образом.

Временной ряд $T$ может быть логически разбит на *сегменты* – непересекающиеся подпоследовательности заданной длины $m$. Здесь и далее без существенного ограничения общности мы можем считать, что $n$ кратно $m$, поскольку $m \ll n$. Множество сегментов ряда, имеющих длину $m \ll n$, обозначим как $S_T^m$, элементы этого множества как $S_1, \ldots, S_{n/m}$:

$$S_T^m = (S_1, \ldots, S_{n/m}), \text{где } S_i = T_{m \cdot (i-1)+1, m}. \qquad (4)$$

Обозначим множество сниппетов ряда $T$, имеющих длину $m \ll n$, как $C_T^m$, а элементы этого множества – как $C_1, \ldots, C_K$:

$$C_T^m = (C_1, \ldots, C_K), \text{где } C_i \in S_T^m. \qquad (5)$$

Число $K$ ($1 \leq K \leq n/m$) представляет собой параметр, задаваемый прикладным программистом, и отражает соответствующее количество наиболее типичных сниппетов. С каждым сниппетом ассоциированы следующие атрибуты: индекс сниппета, ближайшие соседи и значимость данного сниппета.

*Индекс сниппета* $C_i \in C_T^m$ обозначается как $C_i.index$ и представляет собой номер $j$ сегмента $S_j$, которому соответствует подпоследовательность ряда $T_{m \cdot (j-1)+1, m}$.

*Множество ближайших соседей сниппета* $C_i \in C_T^m$ обозначается как $C_i.Neighbors$ и содержит подпоследовательности ряда, которые более похожи на данный сниппет, чем на другие сегменты ряда, в смысле меры схожести MPdist [6]:

$$C_i.Neighbors = \{T_{j,m} \mid S_{C_i.index} = \arg\min_{1 \le r \le n/m} \text{MPdist}(T_{j,m}, S_r), 1 \le j \le n - m + 1\}. \quad (6)$$

*Значимость сниппета* $C_i \in C_T^m$ обозначается как $C_i.frac$ представляет собой долю множества ближайших соседей сниппета в общем количестве подпоследовательностей ряда, имеющих длину $m$:

$$C_i.frac = \frac{|C_i.Neighbors|}{n-m+1}. \quad (7)$$

Сниппеты упорядочиваются по убыванию их значимости:

$$\forall C_i, C_j \in C_T^m: i < j \Leftrightarrow C_i.frac \ge C_j.frac. \quad (8)$$

Таким образом, рассматриваемая задача состоит в том, чтобы для заданных временного ряда $T$, длины сегмента $m$ и количества значимых сниппетов $K$ найти наиболее значимые сниппеты $C_1, \ldots, C_K \in C_T^m$, в т.ч. индекс, множество ближайших соседей и значимость каждого из упомянутых сниппетов.

## 4 Параллельный алгоритм для автоматизированной разметки больших временных рядов

Разработанный алгоритм PaSTiLa (Parallel Snippet-based Time series Labeling) позволяет выполнять поиск сниппетов различных длин на вычислительном кластере с узлами на базе графических процессоров в большом временном ряде (который не может быть целиком размещен в оперативной памяти). Алгоритм выполняет автоматический подбор наилучшего значения длины сниппета из заданного диапазона, что позволяет добиться высокой точности поиска шаблонов и разметки активностей субъекта, деятельность которого описывает данный временной ряд.

### 4.1 Параллельный алгоритм PSF

Для поиска подпоследовательностей-сниппетов в алгоритме PaSTiLa используется модификация разработанного ранее параллельного алгоритма PSF (Parallel Snippet Finder). Вычислительная схема алгоритма PSF имеет следующие отличия от оригинального алгоритма SnippetFinder [6]. Вместо одного последовательного шага, на котором выполняется вычисление MPdist-профиля между сегментом и каждой подпоследовательностью исходного ряда, мы можем выполнить последовательность параллельных шагов.

Ключевой для распараллеливания структурой данных является матрица $\text{ED}_{\text{norm}}$-расстояний между каждой подпоследовательностью длины $\ell$ сегмента $seg$ и каждой подпоследовательностью длины $\ell$ исходного ряда. Параллелизм вычислений матрицы расстояний в алгоритме реализован на основе техники, предложенной в работе [7], что позволяет использовать меньше вычислительных ресурсов по сравнению с прямолинейным выполнением z-нормализации подпоследовательностей и вычислением евклидова расстояния между ними. Обозначим указанную матрицу за $ED_{matr}$:

$$ED_{matr} \in \mathbb{R}^{(m-\ell+1)\times(n-\ell+1)} : ED_{matr}(i,j) = \text{ED}_{\text{norm}}(seg_{i,\ell}, T_{j,\ell}), \qquad (9)$$

На втором шаге в каждом столбце матрицы $ED_{matr}$, полученной на первом шаге, находится минимум. Обозначим вектор таких минимумов за $allP_{BA}$:

$$allP_{BA} \in \mathbb{R}^{n-\ell+1} : allP_{BA}(j) = \min_{1 \leq i \leq m-\ell+1} ED_{matr}(i,j). \qquad (10)$$

На третьем шаге в каждой строке $ED_{matr}$ выполняется поиск минимумов в скользящем окне длины $\ell$. Обозначим матрицу таких минимумов за $allP_{AB}$:

$$allP_{AB} \in \mathbb{R}^{(m-\ell+1)\times(n-\ell+1)} : allP_{AB}(i,j) = \min_{j \leq c \leq j+m-\ell+1} ED_{matr}(i,c). \qquad (11)$$

На четвертом шаге для каждой подпоследовательности ряда, имеющей длину $\ell$, и сегмента $seg$ выполняется построение матричного профиля. Для построения одного матричного профиля выполняется сцепление соответствующих данной подпоследовательности столбца матрицы $allP_{AB}$ и подпоследовательности длины $m - \ell + 1$, входящей в вектор $allP_{BA}$. Результат сцепления обозначим как вектор $P_{ABBA}$:

$$P_{ABBA} \in \mathbb{R}^{2(m-\ell+1)} : P_{ABBA}(T_{j,\ell}) = \{allP_{AB}(i,j)\}_{i=1}^{m-\ell+1} \cdot \{allP_{BA}(i)\}_{i=j}^{m-\ell+1}, \qquad (12)$$

где $1 \leq j \leq n - \ell + 1$.

Далее, обозначим за $SortedP_{ABBA}$ ряд $P_{ABBA}$, в котором элементы упорядочены по возрастанию. Для вычисления схожести между рядами $A$ и $B$ в смысле меры MPdist используется $k$-е значение ряда $SortedP_{ABBA}$, где $k$ является параметром. Типичным значением указанного параметра является 5% от $2m$. Однако, если величина значимой длины подпоследовательности $\ell$ существенно близка к длине ряда $m$, то конкатенация матричных профилей $P_{ABBA}$ имеет длину менее 5% от $2m$. В этом случае значением меры MPdist полагается максимальное значение элемента в конкатенации матричных профилей $P_{ABBA}$. Нижеследующая формула формализует приведенные рассуждения:

$$\text{MPdist}(A,B) = \begin{cases} SortedP_{ABBA}(k) & |P_{ABBA} > k| \\ SortedP_{ABBA}(2(m-\ell-1)) & |P_{ABBA} \leq k| \end{cases}. \qquad (13)$$

MPdist-профилем временного ряда $T$ длины $n$ и временного ряда $Q$ меньшей длины $m$ ($m \ll n$) назовем вектор из $n - m + 1$ элементов, каждый из ко-торых представляет собой величину схожести соответствующей подпоследовательности ряда $T$, имеющей длину $m$, с рядом $Q$, также имеющим длину $m$, в смысле меры MPdist, и обозначим его как $MPD$:

$$MPD(Q,T) = (d_1, \ldots, d_{n-m+1}), d_i = \text{MPdist}(Q, T_{i,m}). \qquad (14)$$

Обозначим MPdist-профиль ряда $T$ и его сегмента $S_i$ как $D_i$, тогда набор MPdist-профилей ряда $T$ и всех его сегментов обозначим как $D$ и определим следующим образом:

$$D = (D_1, \ldots, D_{n/m}), D_i = MPD(S_i, T). \qquad (15)$$

Поиск сниппетов связан с построением кривой репрезентативности, которая состоит из $n - m$ точек и обозначается как $M$. Параметром построения данной кривой является $D_{subset}$, заданное непустое подмножество набора MPdist-профилей. Кривая $M$ представляет собой набор точек, в котором $i$-я точка показывает схожесть по мере MPdist между $i$-й подпоследовательностью ряда, имеющей длину $m$, и наиболее похожим на нее сегментом ряда, взятым из заданного подмножества сегментов:

$$M(D_{subset}) = (M_1, \ldots, M_{m-n+1}), \qquad (16)$$
$$M_i = \min_{D_j \in D_{subset}} \{d_i \mid d_i \in D_j\}, D_{subset} \subset D, \ |D_{subset}| = K.$$

Площадь под кривой репрезентативности $M$ обозначается как ProfileArea и рассматривается как целевая функция поиска сниппетов:
$$ProfileArea(D_{subset}) = \sum_{i=1}^{n-m} M_i(D_{subset}). \qquad (17)$$

Поиск сниппетов подразумевает такой выбор сокращенного множества сегментов ряда в качестве сниппетов, при котором значение $ProfileArea$ существенно приблизится к нулю. По итогу поиска множество $D_{subset}$ будет содержать MPdist-профили K сниппетов, имеющих наибольшую значимость.
Была сделана модификация алгоритма PSF, позволяющая ему задействовать несколько графических ускорителей на одном узле кластера. Это достигается путем равномерного распределения вычислений MPdist профилей между ГПУ.

## 4.2 Параллельный алгоритм PaSTiLa

Алгоритм PaSTiLa предполагает репликацию временного ряда по дискам узлов вычислительного кластера и обеспечивает два уровня параллелизма по данным. На первом уровне параллелизма диапазон длин сниппетов $m_{min} \ldots m_{max}$ распределяется по узлам кластера для последующих вычислений соответствующих MPdist-профилей временного ряда таким образом, чтобы обеспечить баланс загрузки узлов кластера. MPdist-профиль сниппета представляет собой массив, i-й элемент которого содержит значение MPdist-расстояния от заданного сниппета до i-й подпоследовательности ряда. Второй уровень параллелизма заключается в вычислении MPdist-профиля на нескольких графических процессорах одного узла кластера. Для обменов данными между узлами кластера используется технология MPI. Для организации вычислений на графических процессорах применяется технология CUDA.

Для распределения длин сниппетов $m_{min} \dots m_{max}$ по узлам кластера выполняется предсказание времени выполнения параллельного алгоритма поиска сниппетов для различных значений параметра. Для этого используется модель полиномиальной регрессии, которая обучается на основе данных о времени выполнения алгоритма, полученных при его предыдущих запусках. Далее с помощью известного алгоритма оптимизации Кармаркара–Карпа [12] выполняется распределение значений длин сниппетов между узлами кластера, которое обеспечивает сбалансированную вычислительную нагрузку всех узлов. Выполняющийся на узлах кластера параллельный алгоритм поиска сниппетов является модификацией разработанного ранее алгоритма PSF (Parallel Snippet-Finder) [8] и обеспечивает вычисления нескольких графических процессорах узла.

Найденные сниппеты, имеющие различную длину, пересылаются на один узел, после чего выполняется подбор наилучшего значения длины (которая с большей вероятностью даст в итоге наиболее точную разметку активностей субъекта в заданном ряде). Для этого предложен новый критерий выбора длины сниппета. Неформально его можно определить следующим образом: данный критерий отдает предпочтение такой длине сниппета, которая дает в итоге сниппеты, существенно отличающиеся между собой по форме. Формально критерий можно определить следующим образом:

$$m_{best} = \arg \max_{m_{\min} \leq m \leq m_{\min}} \frac{\sum_{\substack{1 \leq p,q \leq K \\ p \neq q}} \sum_{i=1}^{n-m+1} |D_p(i) - D_q(i)|}{\max_{S \in S_T^m} MPD(S,T)}. \qquad (18)$$

наилучшим значением длины сниппетов будет то, которое дает максимальную площадь между кривыми MPdist-профилей этих сниппетов

В результате работы алгоритм PaSTiLa выдает наилучшее значение длины сниппета $m_{best}$ и сниппеты это длины. В отличие от разработанного ранее алгоритма PSF и аналогов, алгоритм PaSTiLa самостоятельно выполняет подбор наилучшего значения параметра, т.е. не требует тонкой настройки. Это позволяет исследователю получать адекватные результаты поиска шаблонов ряда за минимальное время.

## 5 Вычислительные эксперименты

## 5.1 Описание экспериментов

Для исследования эффективности разработанного алгоритма PaSTiLa были проведены вычислительные эксперименты. В экспериментах изучалась масштабируемость алгоритма – его способность пропорционально увеличивать ускорение и производительность при добавлении аппаратных ресурсов (графических процессоров и узлов кластера), и точность разметки временных рядов.

### 5.1.1 Наборы данных и аналоги

Для экспериментов использовались временные ряды из набора данных Time Series Segmentation Benchmark (TSSB) [9], это набор содержит 75 временных рядов из различных предметных областей. Ряд Solar Power Dataset [10] содержит показания об объеме получаемой солнечной энергии, показания мощности записывались каждые 4 секунды, начиная с 01.08.2019. Набор данных PAMAP [11] содержит показания носимого акселерометра во время различных физических активностей человека.

В экспериментах мы проводили сравнение со следующими передовыми алгоритмами сегментации временных рядов: ClaSP [9], FLOSS [13], BinSeg [14]. В качестве входного параметра для алгоритма FLOSS была использована наилучшая длина сниппета, подобранная алгоритмом PaSTiLa. На вход алгоритма BinSeg подавалось количество переходов между активностями ряда. ClasSP выполняет автоматический подбор параметров.

### 5.1.2 Аппаратная платформа

Вычислительные эксперименты были проведены на платформе суперкомпьютера Лобачевский (ННГУ, Нижний Новгород) для следующей конфигурации вычислительного кластера с узлами на базе графических процессоров.

*Конфигурация* $64 \times K20$ использовала 64 графических процессора узлов сегмента суперкомпьютера Лобачевский. На всех узлах установлено по три графических процессора NVIDIA Kepler K20X, каждый из которых имеет следующие характеристики: 2 688 ядер, частота ядра @732 МГц, пиковая производительность 1.31 TFLOPS для арифметики двойной точности.

## 5.2 Результаты

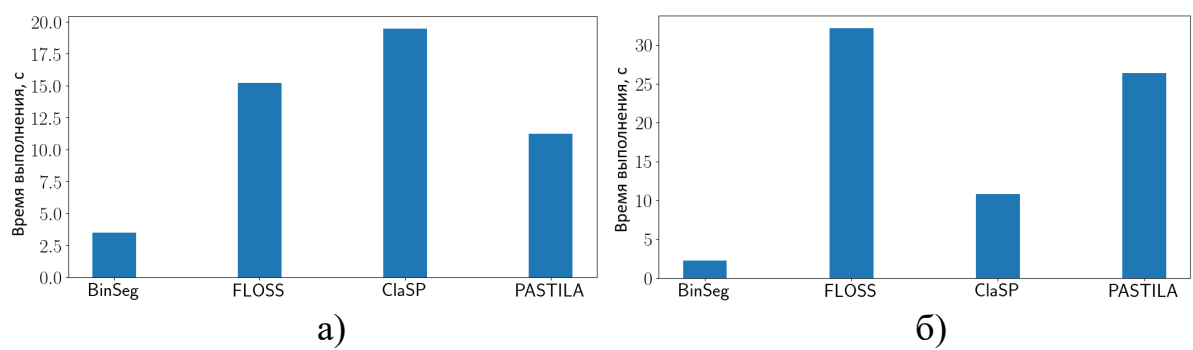

а) б)

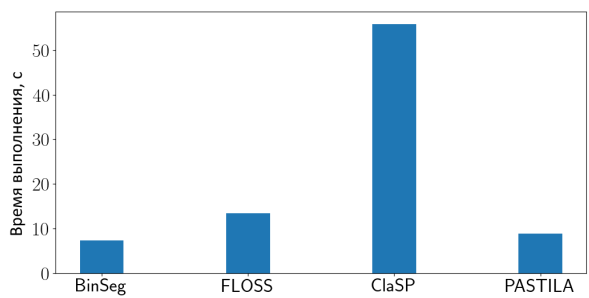

в)

Рис. 5.2.1. Время выполнения алгоритма PaSTiLa на наборе данных TSSB
а) для рядов длины $n < 4000$, б) для рядов длины $4000 \leq n < 8000$,
в) для рядов длины $n \geq 8000$

Графики на рис. 5.2.1 отражают суммарное время работы алгоритма для временных рядов различной длины из набора данных TSSB. Можно видеть, что алгоритм показывает высокую производительность для рядов большей длины ($n \geq 8000$). Понижение производительности на небольших рядах $n < 4000$ можно объяснять тем, что время, затрачиваемое на пересылку данных, становится соразмерным с основными вычислениями.

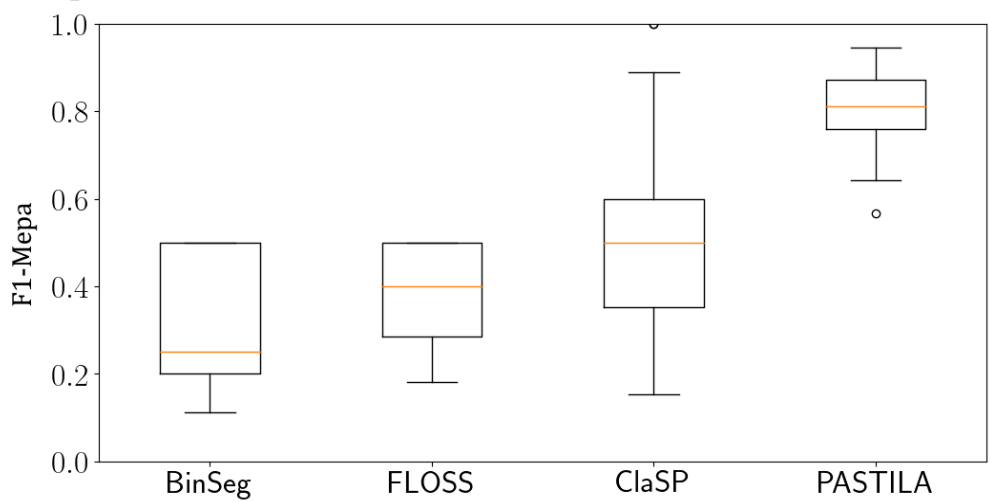

Рис. 5.2.2. Точность работы алгоритма PaSTiLa на наборе данных TSSB

Для оценки точности разметки временных рядов нами использованы стандартные мера качества классификации F1-мера, определяемая следующим образом:

$$Precision = \frac{TP}{TP + FP}, Recall = \frac{TP}{TP + FN}, F1 = 2 \cdot \frac{Precision \cdot Recall}{Precision + Recall},$$

где $TP$, $FP$, $TN$ и $FN$ — количество истинно-положительных, ложно-положительных, истинно-отрицательных и ложно-отрицательных элементов ряда соответственно при сравнении истинной и полученной при помощи алгоритма разметок ряда. На рис. 5.2.2 видно, что алгоритм PaSTiLa выполняет разметку временных рядов со значительно большей точность по сравнению с аналогами.

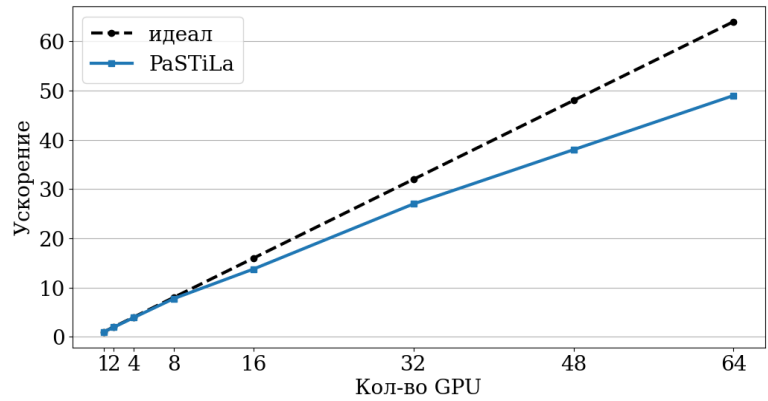

Рис. 5.2.3. Ускорение алгоритма PaSTiLa

На рис. 5.2.3 можно видеть результаты экспериментов по оценке ускорения алгоритма во временном ряду Solar Power Dataset. На основе полученных результатов можно видеть, что при количестве GPU менее 32 ускорение близко к линейному

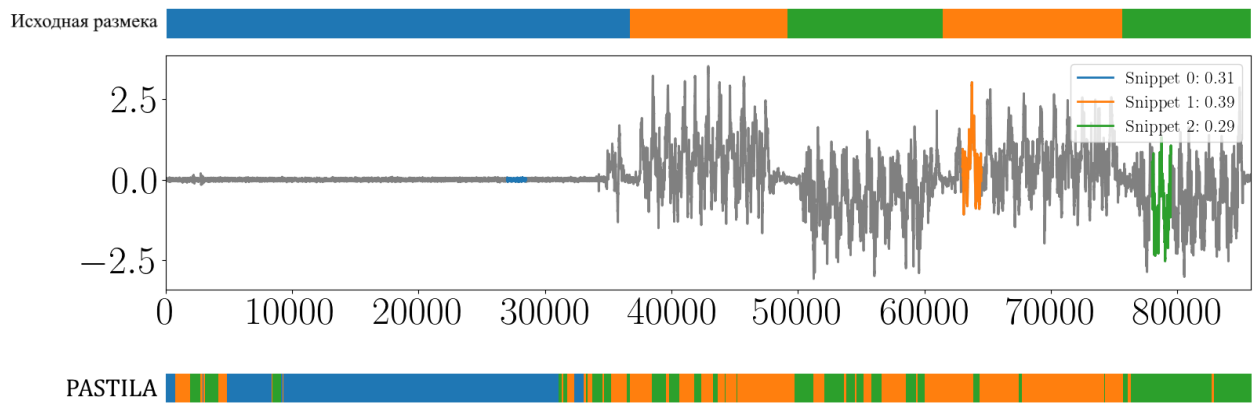

Рис. 5.2.4. Разметка временного ряда PAMAP при помощи алгоритма PaSTiLa

Были проведены тематические исследования с использованием набора данных PAMAP. На рис. 5.2.4 можно видеть результат поиска сниппетов и разметки временного ряда с трема активностями: состояние покоя, подъем по лестнице и спуск по лестнице. Разработанный алгоритм успешно выявил сниппеты временного ряда и дал адекватную разметку присутствующих активностей.

# 6 Заключение

Статья посвящена проблеме применения параллельных вычислений на графическом процессоре для повышения производительности и качества поиска шаблонов временных рядов. Поиск шаблонов предполагает автоматизированный поиск наиболее типичных подпоследовательностей, которые выделяют различные активности, заданные указанным рядом. Поиск шаблонов имеет широкий спектр практического применения: в приложениях цифровой индустрии, интеллектуальное управление зданиями в приложениях Интернета вещей, мониторинг состояния человека и упреждающая диагностика заболеваний в приложениях персональной медицины и др.

Разработанный алгоритм PaSTiLa позволяет выполнять поиск шаблонов различной длины в больших временных рядах. Был предложен критерий для выбора наилучшей длины шаблона, что позволяет максимизировать точность поиска, исключая возможность неправильного выбора значения параметра. Алгоритм имеет высокую масштабируемость и может быть эффективно запущен на вычислительных кластерах с большим количеством узлов.

Результаты экспериментов подтверждают высокую эффективность алгоритма PaSTiLA и демонстрируют, что предложенный алгоритм имеет более высокую производительность и точность поиска шаблонов по сравнению с аналогами.

# Список литературы

1  Imani S., Madrid F., Ding W., Crouter S., Keogh E. Matrix Profile XIII: Time Series Snippets: A New Primitive for Time Series Data Mining // Proc. of the 9th IEEE Int. Conf. on Big Knowledge, IEEE ICBK 2018 (Singapore, November 17 – November 18, 2018). IEEE Press, 2018. P. 383–389. DOI: 10.1109/icbk.2018.00058.

2  Chiu, B.Y.; Keogh, E.J.; Lonardi, S. Probabilistic discovery of time series motifs. In Proceedings of the 9th ACM SIGKDD International Conference on Knowledge Discovery and Data Mining, Washington, DC, USA, 24–27 August 2003; Getoor, L., Senator, T.E., Domingos, P.M., Faloutsos, C., Eds.; ACM: New York, NY, USA, 2003; pp. 493–498. DOI: 10.1145/956750.956808.

3  Keogh, E.J.; Lin, J. Clustering of time-series subsequences is meaningless: implications for previous and future research. Knowl. Inf. Syst. 2005, 8, 154–177. DOI: 10.1007/s10115-004-0172-7.

4  Mueen A., Nath S., Liu J. Fast approximate correlation for massive time-series data // Proc. of the ACM SIGMOD Int. Conf. on Management of Data, SIGMOD 2010, Indianapolis, Indiana, USA, June 6-10, 2010. P. 171–182. DOI: 10.1145/1807167.1807188.

5  Yankov D., Keogh E.J., Rebbapragada U. Disk aware discord discovery: finding unusual time series in terabyte sized datasets // Knowl. Inf. Syst. 2008. Vol. 17, no. 2. P. 241–262. DOI: 10.1007/s10115-008-0131-9.

6  Gharghabi S., Imani S., Bagnall A., Darvishzadeh A., Keogh E. An Ultra-Fast Time Series Distance Measure to allow Data Mining in more Complex Real-World Deployments // Data Mining and Knowledge Discovery. Vol. 34. Berlin: Springer, 2020. P. 1104–1135. DOI: 10.1007/s10618-020-00695-8.

7  Yeh C.M., Zhu Y., Ulanova L., et al. Time series joins, motifs, discords and shapelets: a unifying view that exploits the matrix profile // Data Mining and Knowledge Discovery, 2018. Vol. 32, no. 1. P. 83–123. DOI: 10.1007/s10618-017-0519-9.

8  Цымблер М.Л., Гоглачев А.И. Поиск типичных подпоследовательностей временного ряда на графическом процессоре // Вычислительные методы и программирование. 2021. № 4. С. 344–359. DOI: 10.26089/NumMet.v22r423.

9  Ermshaus A., Schäfer P., Leser U. ClaSP: parameter-free time series segmentation. Data Min. Knowl. Discov. 37(3): P. 1262-1300, 2023. DOI: 10.1007/s10618-023-00923-x

10 Godahewa R., Bergmeir C., Webb G., Abolghasemi M., Hyndman R., Montero-Manso P. Solar Power Dataset (4 Seconds Observations). https://zenodo.org/records/4656027

11 Reiss A., Stricker D. Introducing a New Benchmarked Dataset for Activity Monitoring // Proceedings of the 16th International Symposium on Wearable Computers, ISWC 2012 (Newcastle, United Kingdom, June 18-22, 2012). P. 108-109. DOI:10.1109/iswc.2012.13.

12 Karmarkar N., Karp R. The differencing method of set partitioning, Tech report UCB/CSD 82/113, Computer science division, University of California, Berkeley, 1982

| 13 | Gharghabi, S., Yeh, C.M., Ding, Y., Ding, W., Hibbing, P., LaMunion, S., Kaplan, A., Crouter, S.E., Keogh, E.J.: Domain agnostic online semantic segmentation for multi-dimensional time series. Data Min. Knowl. Discov. 33(1), P. 96–130, 2019. DOI: 10.1007/S10618-018-0589-3 |
|----|---|
| 14 | Truong, C., Oudre, L., Vayatis, N.: Selective review of offline change point detection methods. Signal Process. 167 (2020). DOI: 10.1016/J.SIGPRO.2019.10729 |

**Информация об авторах**

*Гоглачев Андрей Игоревич* — программист лаборатории суперкомпьютерного моделирования; Южно-Уральский государственный университет (национальный исследовательский университет), пр-т им. В. И. Ленина, д. 76, 454080, Челябинск, Российская Федерация